\documentclass[12pt]{article}
\usepackage[T1]{fontenc}
\usepackage[utf8]{inputenc}
\usepackage[margin=1in]{geometry}
\usepackage[affil-it]{authblk} 
\usepackage{etoolbox}
\usepackage{lmodern}
\usepackage{cite}
\usepackage{changes}
\usepackage{microtype}
\usepackage{booktabs}
\usepackage{amssymb}
\usepackage{float}
\DeclareUnicodeCharacter{030C}{\v{c}}

\title{Triangular Cross-Section Beam Splitters in Silicon Carbide for Quantum Information Processing}

\def\correspondingauthor{\footnote{Corresponding author: smajety@ucdavis.edu}}

\author[1] {Sridhar Majety \correspondingauthor{}\textsuperscript{†}}
\author[1] {Pranta Saha \textsuperscript{†}}
\author[1] {Zbynka Kekula}
\author[2] {Scott Dhuey}
\author[1] {Marina Radulaski}

\affil[1]{Department of Electrical and Computer Engineering, University of California, Davis, CA 95616, USA}
\affil[2]{The Molecular Foundry, Lawrence Berkeley National Laboratory, Berkeley, CA 94720, USA}

\date{\vspace{-2em}}

\usepackage{graphicx}
\usepackage{amsmath}
\usepackage{braket}
\usepackage{titlesec}
\usepackage{siunitx}
\usepackage[version=4]{mhchem}
\usepackage{multirow}
\begin{document}

\maketitle
\vspace{-0.8cm}

\let\thefootnote\relax\footnotetext{\textsuperscript{†} These authors contributed equally to this work}
\begin{abstract}
Triangular cross-section color center photonics in silicon carbide is a leading candidate for scalable implementation of quantum hardware. Within this geometry, we model low-loss beam splitters for applications in key quantum optical operations such as entanglement and single-photon interferometry. We consider triangular cross-section single-mode waveguides for the design of a directional coupler. We optimize parameters for a 50:50 beam splitter. Finally, we test the experimental feasibility of the designs by fabricating triangular waveguides in an ion beam etching process and identify suitable designs for short-term implementation.

\end{abstract}

\section{Introduction}
Photonic quantum information processing (QIP) using solid-state single photon emitters has been extensively explored for applications in quantum communications and integrated quantum information circuits.  
Among the quantum emitters, color centers have been desired solid state qubit candidates due to their spectral homogeneity, long spin coherence times, and availability of spin-spin and spin-photon entangling processes, necessary for QIP applications \cite{norman2021novel, bathen2021manipulating, son2020developing, castelletto2020silicon, lukin2020integrated, majety2022quantum}. 
Particularly, color centers in silicon carbide (SiC) attracted attention because of emissions in the telecommunication bands which are suitable for sending over long distances in fiber optic cable. 
SiC offers other advantages like large bandgap, strong second-order non-linearity, decades-long industry presence, and is CMOS compatible.

The existing QIP algorithms require error correction and heralded measurements to process information reliably and efficiently. Hence, one of the near-term goals is to build the on-chip programmable photonic mesh networks to perform fault-tolerant QIP \cite{hamerly2022asymptotically, dong2023programmable, bogaerts2020programmable}. Successful implementation of such quantum mesh photonic hardware requires efficient single photon generation, manipulation, entanglement, and detection \cite{gyger2021reconfigurable, martini2019single, majety2023triangular}. Beamsplitters (BS) play a crucial role in on-chip photonic QIP circuits for generating multiphoton correlation \cite{reck1994experimental, dong2023programmable} and can be used to realize several important operations like entanglement generation and single-photon interferometry (Hong-Ou-Mandel and Hanbury Brown-Twiss). Moreover, a universal quantum computer with linear optical quantum computing protocols has been proposed using BS, phase shifters, single photon sources, and photo-detectors \cite{knill2001scheme}. Among the most studied forms of BS such as Y-branch, directional coupler (DC), and multi mode  interferometer (MMI), DC is the most suited for on-chip QIP as it can function with single-mode waveguides \cite{xu2022methods}. 

Color centers properties are preserved in the bulk of the host material and for successful nanophotonic integration of these color centers, it is necessary to be able to grow or attach a high quality thin film on a substrate with high refractive index contrast in a scalable way. Triangular photonics has been the top choice in diamond, silicon carbide and rare earth photonics \cite{burek2012free, kindem2020control, babin2022fabrication} that provides pristine color centers and high light and matter interaction. Compared to the other non-traditional processing techniques \cite{ferro20153c, bracher2017selective, lukin20204h} used for fabricating undercut devices, angle etching method \cite{burek2012free, song2018high, babin2022fabrication} which produces triangular cross-section and undercut photonic devices, is promising for wafer-scale production \cite{atikian2017freestanding}. Performance of a variety of such triangular cross-section photonic structures in 4H-SiC were recently examined \cite{babin2022fabrication, majety2021quantum, majety2023triangular, saha2023utilizing}, and have proven favorable for QIP applications.

In this paper, we explore triangular cross-section 50:50 BS in 4H-SiC necessary for performing on-chip quantum interferometry.
We construct such a device that splits an input beam into two output beams of equal intensity using two concurrent waveguides. In the front and back regions, the waveguides are far apart (no electromagnetic interaction) on either ends (input, output) of the device, while in the middle overlap region waveguides propagate in close proximity to create an electromagnetic interaction between their fields. When two indistinguishable photons each entangled to a spin of a quantum emitter arrive at the inputs of a 50:50 BS at the same time, they undergo bunching to exit together through the same output, creating remote entanglement between the two spins. Here we specifically study 50:50 BS designs for nitrogen vacancy (NV) color centers in 4H-SiC, with single photon emissions in the telecommunication wavelength range (1176 - 1243 nm) \cite{mu2020coherent, sato2019formation}.  

Using Finite-Difference Time-Domain (FDTD) method, we first optimize the width of triangular cross-section waveguides to achieve single mode propagation (fundamental TE mode), and high coupling efficiency of the color center (NV in 4H-SiC) emission into that single mode, for different etch angles. Next, we investigate the formation of supermodes when two identical single mode waveguides are placed adjacent to each other. We then study the waveguide mode conversion in bent waveguides and find suitable waveguide bend geometries to maintain the same waveguide mode in the straight and bent regions of the waveguide. We use this understanding to design triangular cross-section 50:50 BS in 4H-SiC. Finally, we explore the fabrication of the simulated triangular cross-section waveguide configurations required for a 50:50 BS.    

\section{Single mode triangular cross-section waveguide}
Color center emission is dipole-like and such emission can couple to fundamental TE (f-TE), fundamental TM (f-TM) and other higher order modes that are supported in triangular cross-section waveguides \cite{majety2021quantum}. It has been shown that for each etch angle there exists an optimal width for single mode (f-TE) propagation in triangular cross-section waveguides \cite{babin2022fabrication}, which is a necessity for QIP applications \cite{caves1994quantum}. So, to achieve high coupling efficiencies for color center emission, the color center should be positioned in a single mode waveguide at the maximum electric field (E-field) intensity point of the f-TE mode\cite{babin2022fabrication, majety2023triangular}, which is at the centroid of the triangle. 

\begin{figure}[!hb]
    \centering
    \includegraphics[width=\textwidth]{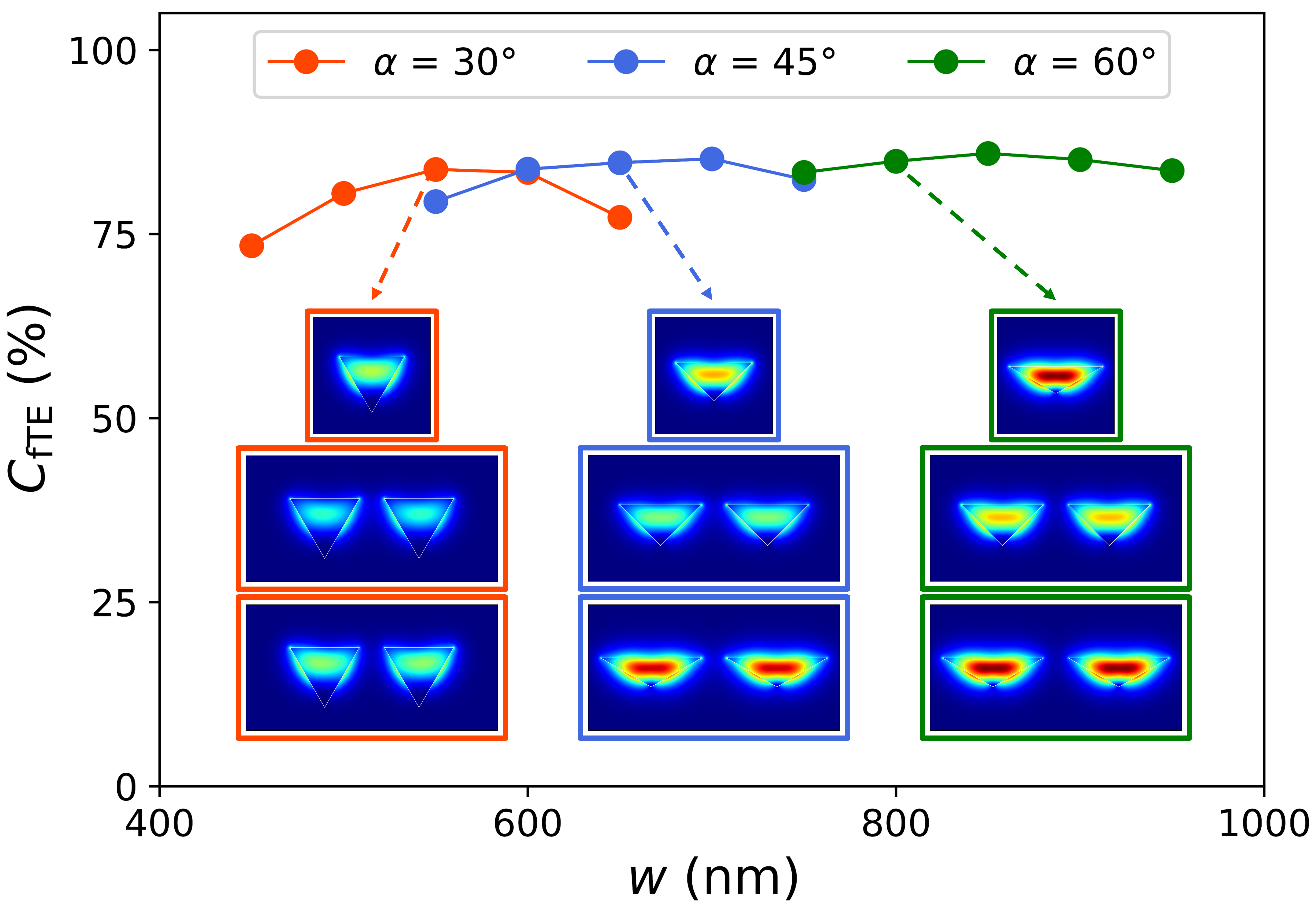}
    \caption{The coupling of a TE oriented dipole emission at the centroid of the triangular profile into the f-TE mode (($C_\mathrm{fTE}$)) as a function of the waveguide width for three etch angles ($\alpha$). Insets show electric field intensity profiles of the f-TE mode, TE supermode 1 and TE supermode 2 (top to bottom), at optimal waveguide widths supporting single mode propagation, for $\alpha = 30^{\circ}, 45^{\circ}, 60^{\circ}$. The electric field intensity of the TE supermodes were plotted for adjacent identical waveguides with 200 nm gap between them.}
    \label{fig:fig_1}
\end{figure}

We use FDTD package in Lumerical software to estimate the coupling efficiency of color center emission into the f-TE mode ($C_\mathrm{fTE}$) for different widths ($w$) and etch angles ($\alpha$) of a triangular cross-section waveguide (the angle at the apex of the triangle is $2\alpha$). We position the dipole emitter at the centroid of the triangle with an emission wavelength of 1230 nm. We find that for each $\alpha$ there exists a width with highest coupling ($>80\%$) to the f-TE mode as shown in Figure \ref{fig:fig_1}. For our 50:50 BS simulations, we choose the optimal width values ($w$ = 550 nm, 650 nm, 800 nm for $\alpha = 30^{\circ}, 45^{\circ}, 60^{\circ}$ respectively) slightly lower than the widths with the highest $C_\mathrm{fTE}$. The reason being that the waveguides with lower width have slightly higher evanescent fields (top panels of inset in Figure \ref{fig:fig_1}), necessary to create meaningful coupling between the adjacent waveguides in the BS, while only slightly reducing $C_\mathrm{fTE}$ ($< 1\%$). 

The mode profile of the waveguide f-TE (f-TM) mode has a E-field intensity maximum closer to the top surface (apex) \cite{majety2021quantum}. As the waveguide width decreases, the E-field intensity maximum of the f-TM mode moves closer to the apex and becomes evanascent (not supported in the waveguide). Using MODE package in Lumerical software, we find that at these optimal widths for $\alpha = 45^{\circ}$ and $60^{\circ}$, the waveguide supports only the f-TE mode. However, for $\alpha = 30^{\circ}$, both f-TE and f-TM modes are supported in the waveguide at the optimal  and smaller widths. The height of the triangle varies inversely with $\alpha$ and hence the widths needed for the f-TM to become evanescent are much smaller than the optimal width for $\alpha = 30^{\circ}$.

\begin{figure}[hb!]
\centering\includegraphics [width=\textwidth]{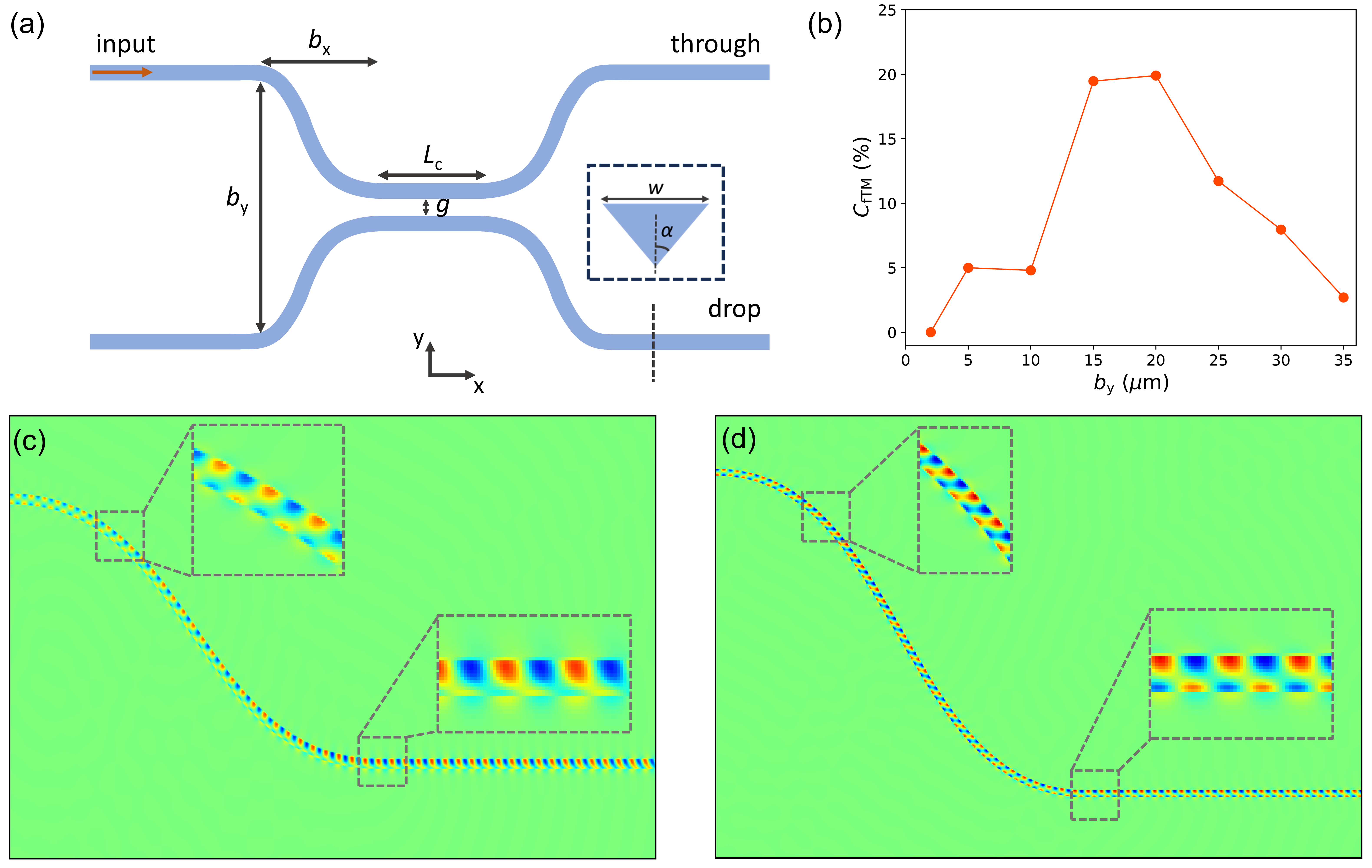}
\caption{(a) Top view of the S-bend BS in 4H-SiC and the inset shows the triangular cross-section geometry. (b) The fraction of the f-TE input mode coupling to f-TM ($C_\mathrm{fTM}$) mode along the bend  with variations in bend y-span ($b_\mathrm{y}$) for $\alpha$ = $30^{\circ}$, $w$ = 550 nm, and $L_\mathrm{C}$ = 30 $\mu$m. (c)-(d) Propagation of $E_\mathrm{z}$ field in the $xy$ plane of the structure described in (b) for $b_\mathrm{y}$ values of 20 $\mu$m and 35 $\mu$m, respectively.}
\label{fig:fig_2}
\end{figure}

In this paper, we design a 50:50 BS using two identical triangular cross-section waveguides with three regions: 1) input and output - where the waveguides are far apart with zero electromagnetic interaction, on either ends of the BS, 2) bent waveguide region, and 3) coupling region - where the waveguides are close enough to  electromagnetically interact, as shown in Figure \ref{fig:fig_2}a. When light is injected into the f-TE mode of one of the waveguides on the input side (top left), it propagates through the straight and bent regions of that waveguide. When it reaches the coupling region, some of the light in the top waveguide is coupled into the bottom waveguide as it propagates along. Then the light in both the waveguides pass through the bent and straight regions to reach the output ports (through and drop ports). The proportion of light coupled into the bottom waveguide is determined by the coupling strength between the individual waveguide modes and the length of the waveguides in the coupling region. 

When two identical waveguides are close to each other, the individual modes (f-TE) in each of the waveguides superimpose to form a supermode. The coupling between the two waveguides can be analyzed in terms of a pair of TE supermodes, as shown in the inset of Figure \ref{fig:fig_1}. For a BS with power $P_\mathrm{0}$ in one of the waveguide at the beginning of the coupling region, the coupling length ($L_\mathrm{C}$) required for a power $P_\mathrm{2}$ to couple into the other waveguide is given by:  
\begin{equation}\label{eq1}
L_\mathrm{C} = \frac{\lambda_0}{\pi \Delta n_\mathrm{eff}} \sin^{-1} \sqrt{\frac {P_2}{P_0}}
\end{equation}

\noindent where $\lambda_0$ is the free space wavelength, $\Delta n_\mathrm{eff}$ is the difference in effective refractive indicies of the two TE supermodes \cite{yuan2010refractive}. The coupling length is inversely proportional to the coupling strength between the waveguides given by $\Delta n_\mathrm{eff}$, which in turn depends on the confinement of the individual waveguide modes and gap between the individual waveguides in the coupling region. For a 50:50 BS, where there is 50\% light in both the waveguides after the coupling region, the ratio $P_\mathrm{2}$/$P_\mathrm{0}$ equals 0.5. We use MODE package in Lumerical software to estimate the effective index of the two TE supermodes, for calculating the theoretical values of $L_\mathrm{C}$ for various BS geometries studied in the following section. 

\section{Integrated 4H-SiC beam splitter in triangular geometry} 
Recent advances in on-chip splitting mostly include rectangular or slab waveguides \cite{xu2022methods}. For performing QIP with integrated color centers in 4H-SiC, we construct the BS structure with symmetric triangular cross-section single mode S-bend waveguides with an overlap region to enable DC. Figure \ref{fig:fig_2}a shows the top view schematic of the BS with coupling length $L_\mathrm{C}$, gap between the waveguides in the coupling region $g$, S-bend x-span $b_\mathrm{x}$, S-bend y-span $b_\mathrm{y}$, and the inset shows the triangular waveguide cross-section with $w$ and $\alpha$. In this paper, we investigate 50:50 BS in triangular cross-section waveguides with $\alpha$ = $30^{\circ}$, $45^{\circ}$, $60^{\circ}$, can be fabricated with the state-of-the-art processes, by varying the gap $g$ between waveguides and consequently $L_\mathrm{C}$.

In Lumerical FDTD simulations (mesh size = 30 nm), we choose 20 $\mu$m for $b_\mathrm{x}$ and 10 $\mu$m for $b_\mathrm{y}$, where $b_\mathrm{x}$ corresponds to the terminal points and $b_\mathrm{y}$ corresponds to the curvature control points of the S-bend (Bézier curve), for $\alpha$ = $45^{\circ}$, $60^{\circ}$. For $\alpha$ = $30^{\circ}$, we need to account for potential higher-order mode conversion due to the bending curvature as the $n_\mathrm{eff}$, one of the most important parameters for understanding mode propagation, of the f-TE and the f-TM modes are virtually the same in the waveguides with $\alpha$ = $30^{\circ}$. Hence, the f-TE mode can transform into the f-TM due to identical $n_\mathrm{eff}$ and inter-mode coupling via bends \cite{zhang2020ultrahigh}. Keeping $b_\mathrm{x}$ = 20 $\mu$m constant, we vary $b_\mathrm{y}$ for modulating the bend curvature of the \{$\alpha$, $w$, $L_\mathrm{C}$\} = \{$30^{\circ}$, 550 nm, 30 $\mu$m\} BS structure. 

To test the inter-mode coupling (cross-talk), we inject via the f-TE mode at the input port and collect via the f-TM mode at the through port of the BS. It is observed in Figure \ref{fig:fig_2}b that coupling to the f-TM mode ($C_\mathrm{fTM}$) initially increases with the increase in $b_\mathrm{y}$ and after reaching a maximum value of $\sim$ 20\% for $b_\mathrm{y}$ = 20 $\mu$m, $C_\mathrm{fTM}$ gradually decreases dropping down to 2.7\% with $b_\mathrm{y}$ = 35 $\mu$m. This trend occurs due to the expansion of the curved path with increasing $b_\mathrm{y}$, however, after the inversion point, the bending becomes so gradual that the S-bend mimics a straight path. For a better understanding of the f-TE and f-TM mode coupling, we also study the $E_\mathrm{z}$ fields of the minimum and maximum $C_\mathrm{fTM}$ as $E_\mathrm{z}$ field is a good indicator of the existence of the TE and TM modes \cite{saha2023utilizing}. Figure \ref{fig:fig_2}c shows that at the starting point of the bending region with $b_\mathrm{y}$ = 20 $\mu$m, $E_\mathrm{z}$ has a nodal plane in the triangular waveguide which corresponds to the f-TE mode and at the ending point,  $E_\mathrm{z}$ field starts to develop in the waveguide, confirming the coupling to the f-TM mode. On the other hand, for $b_\mathrm{y}$ value of 35 $\mu$m, there is a constant $E_\mathrm{z}$ nodal plane indicating the f-TE mode throughout the S-bend, depicted in Figure \ref{fig:fig_2}d, which makes this bend structure suitable for maintaining the single mode propagation in the BS.

\begin{figure}[ht!]
\centering\includegraphics [width=\textwidth]{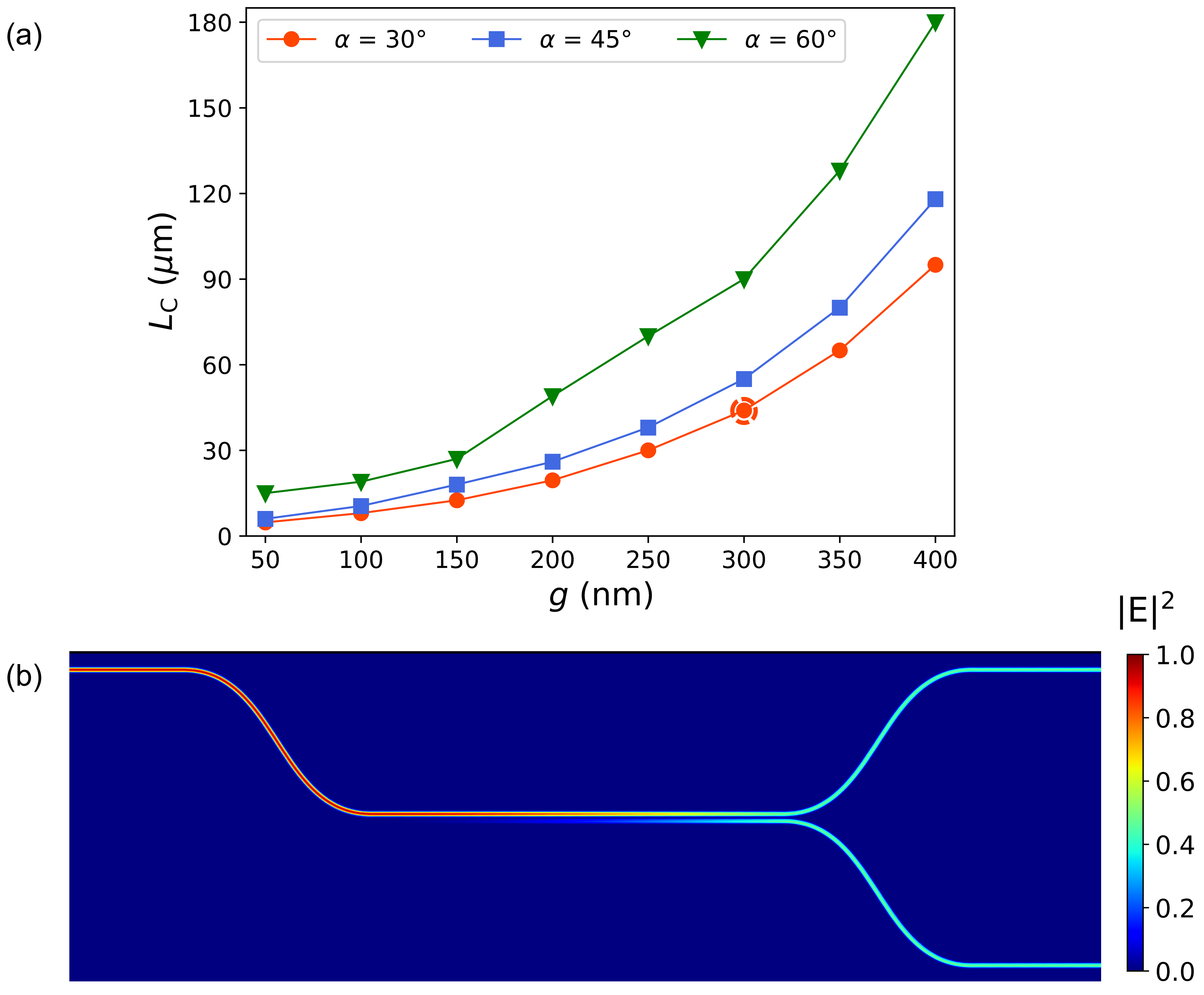}
\caption{(a) $L_\mathrm{C}$ values for 50:50 splitting with an increasing gap $g$ in $\alpha$ =  $30^{\circ}$, $45^{\circ}$, $60^{\circ}$ BS. (b) Top view of the $E$-field intensity profile for the integrated 50:50 BS SiC structure with \{$\alpha$, $w$, $b_\mathrm{x}$, $b_\mathrm{y}$, $g$, $L_\mathrm{C}$\}  = \{$30^{\circ}$, 550 nm, 20 $\mu$m, 35 $\mu$m, 300 nm, 44 $\mu$m\} indicated by dashed circle in Figure \ref{fig:fig_3}a.}
\label{fig:fig_3}
\end{figure}

The above discussions lead us to the optimal selection of $w$ and $b_\mathrm{y}$ of the 50:50 BS for each $\alpha$. We choose 650 nm and 800 nm as $w$ for $\alpha$ = $45^{\circ}$, $60^{\circ}$ respectively with $b_\mathrm{y}$ = 10 $\mu$m in order to maximize the coupling and maintain single mode propagation for establishing a suitable environment for QIP with the triangular cross-section BS. For $\alpha$ = $30^\circ$ waveguides, we pick $w$ = 550 nm and $b_\mathrm{y}$ = 35 $\mu$m to achieve similar outcomes. We vary the gap $g$ from 50 nm to 400 nm for exploring a range of designs. Figure \ref{fig:fig_3}a shows that with increasing $g$, $L_\mathrm{C}$ shoots up rapidly for all three etch angles, experiencing minor losses ($<$ 1.5\%) upon propagation through the entire BS. We observe another interesting phenomenon that $L_\mathrm{C}$ for the 50:50 BS with $\alpha$ = $60^\circ$  is dramatically longer than the $\alpha$ = $30^\circ$ and $\alpha$ = $45^\circ$ BS for a particular $g$ when $g \geq$ 200 nm. As a result, owing to smaller footprints, $\alpha$ = $30^\circ$ and $\alpha$ = $45^\circ$ BS appear as better candidates than the $\alpha$ = $60^\circ$ BS for integrated quantum mesh photonic circuitry.

\section{Fabrication of closely spaced triangular waveguides in 4H-SiC}

Triangular cross-section photonics has previously been fabricated in SiC through Faraday cage assisted etching method \cite{burek2012free, song2018high, babin2022fabrication} suitable for chip-scale integration of color centers with no degradation in color-center properties compared to bulk substrate\cite{babin2022fabrication}. This approach could be brought to wafer-scale using the ion beam etching \cite{atikian2017freestanding}. Therefore, we test a new ion beam etching process to fabricate triangular cross-section waveguides in 4H-SiC \cite{majety2023wafer}, with varying gaps between waveguides (100 nm, 200 nm, 300 nm, 500 nm). This provides an insight into what spacing between the waveguides in the coupling region of the simulated 50:50 BS are fabrication friendly, using the ion beam etch process. 

\begin{figure}[!ht]
    \centering
    \includegraphics[width=\textwidth]{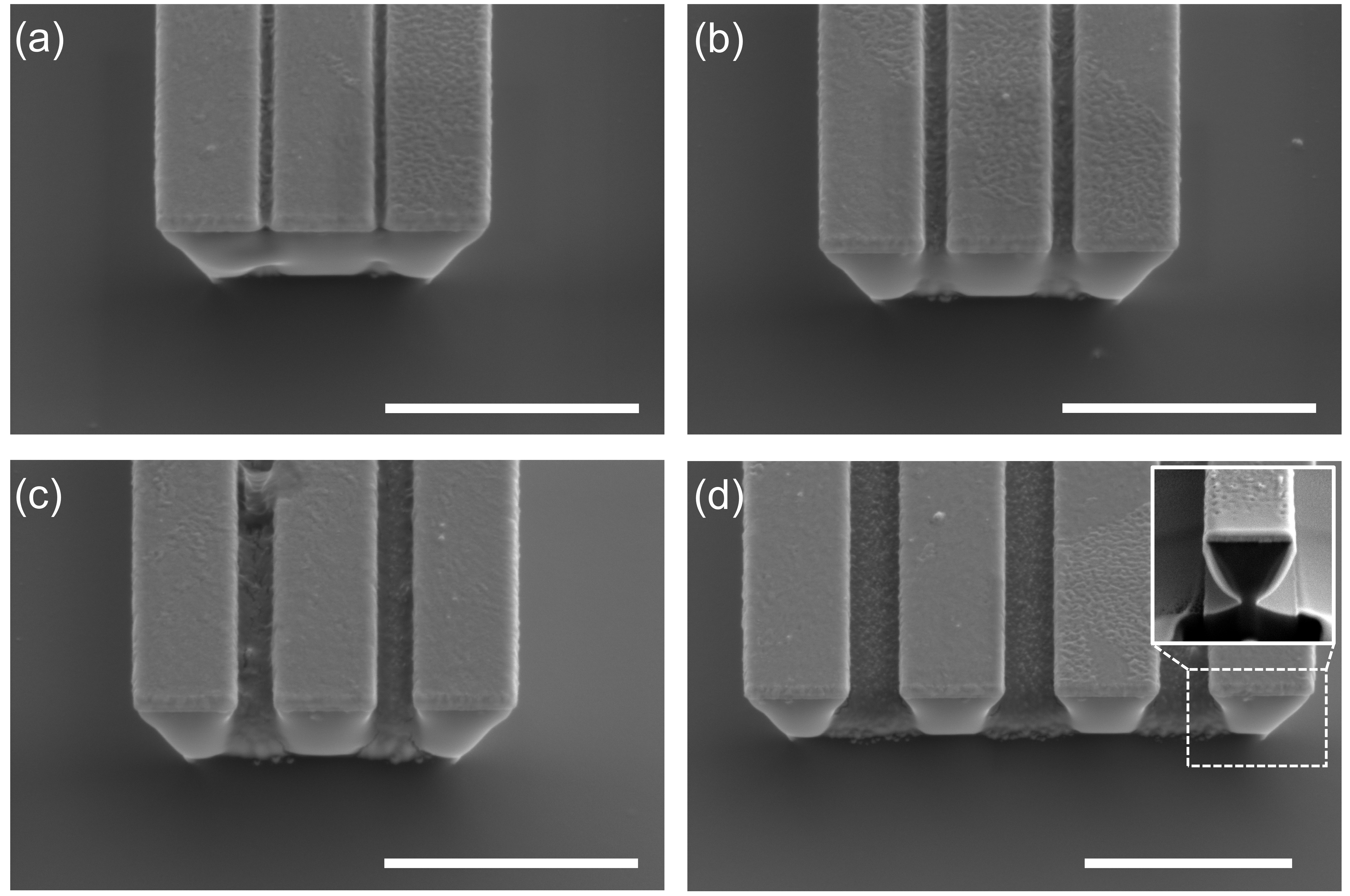}
    \caption{The SEM images of ion beam etched waveguides in 4H-SiC. (a)-(c) 800 nm wide waveguides with gaps of 100 nm, 200 nm and 300 nm respectively. (d) 1000 nm wide waveguides with a gap of 500 nm. Inset shows a FIB-SEM of the fabricated 1000 nm waveguide with a triangular cross-section. The scale bar is 2 $\mu$m.}
    \label{fig:fig_4}
\end{figure}

We use electron beam lithography to define the patterns on 4H-SiC substrates and then the pattern is transferred to a 120 nm thick nickel hard mask deposited using an electron beam evaporation, through a lift-off process. Triangular cross-section waveguides are then fabricated by ion beam etch using SF$_6$ and O$_2$ chemistry. We use focused ion beam (FIB) - scanning electron microscope (SEM) to confirm the triangular cross-section of the fabricated waveguides, as shown in the inset of Figure \ref{fig:fig_4}d. We measure the etch angle ($\alpha$) of these waveguides to be $30^\circ$. For adjacent waveguides with gaps of 100 nm and 200 nm, we observe no substantial etch in the region between the waveguides, as shown in Figure \ref{fig:fig_4}a-b. For waveguide gaps $\geq$ 300 nm (300 nm, 500 nm), the fabricated waveguides have triangular cross-sections, as shown in Figure \ref{fig:fig_4}c-d. The lack of etching in the region between the waveguides for smaller waveguide gaps may be due to the inability of the ions which are incident at an angle on the substrate, to reach well below the surface level of the substrate, and also reduced etching caused by mass-transport issues in those closely spaced regions.

\section{Discussion}
The modeling results presented in this paper offer an approach to build low-loss triangular cross-section 50:50 BS in 4H-SiC necessary to perform key quantum interferometry operations for QIP applications. Here, triangular cross-section photonics provide a scalable route to integrate  color centers into quantum photonic devices and circuits. Moreover, high-performance triangular cross-section photonic structures in 4H-SiC like waveguides, waveguides with integrated SNSPDs, photonic crystal mirrors, and photonic crystal cavities that facilitate efficient generation, collection and detection of single photon emission from color centers, necessary for applications in QIP have been demonstrated \cite{babin2022fabrication, majety2021quantum, majety2023triangular, saha2023utilizing}. Thereby, NV center in 4H-SiC with emission wavelengths near the telecommunications range is most suited for building large-scale quantum communication networks.     

Our initial fabrication tests show that triangular cross-section waveguides with gaps greater than 300 nm can be fabricated using the wafer-scale ion beam etch process. From the simulations, we note that for a waveguide gap $g \geq$ 300 nm in the coupling region, $\alpha$ = $30^\circ$ and $45^\circ$, offer 50:50 BS with fabricatable footprints (e.g. coupling region $L_C \leq$ 50 $\mu$m), for applications in QIP. Here, supporting structures for waveguide suspension would need to be designed. The ion beam etch process conditions could be further optimized to achieve triangular cross-section waveguides with waveguide gap $\leq$ 300 nm, by choosing process conditions that result in a predominantly chemical etch. Under such etch conditions, the byproducts of the etch are volatile gases (SiF$_4$, CO, CO$_2$), preventing re-deposition, allowing the etch to continue. When there is a reasonable physical etch component, the sputtered material has a higher chance of re-deposition, especially when the gap between the waveguides is small, resulting in a slower etch or sometimes no etch. Another alternative could be choosing a steeper etch angle, because such waveguides would not require a deep etch to release the structure.

Triangular cross-section photonics in 4H-SiC provides an avenue to achieve efficient chip- and wafer-scale integration of color centers, with very little degradation of color center properties compared to bulk, essential for applications in QIP. Our simulation results demonstrate that 50:50 BS necessary for applications in QIP can be implemented using triangular geometry and our initial etch tests show that ion beam etching is suitable for achieving this.

\section{Acknowledgments}

M.R., S.M. and P.S. acknowledge support from NSF CAREER (Award 2047564) and AFOSR Young Investigator Program (Award FA9550-23-1-0266). Z.K. acknowledges support by CITRIS Workforce Innovation Program. Work at the Molecular Foundry was supported by the Office of Science, Office of Basic Energy Sciences, of the U.S. Department of Energy under Contract No. DE-AC02-05CH11231. Part of this study was carried out at the UC Davis Center for Nano and Micro Manufacturing (CNM2).

\bibliographystyle{unsrt}
\bibliography{References.bib}

\end{document}